\documentclass{elsart}
\usepackage{epsf}

\newcommand{\MKL}{M_{K_L}}
\newcommand{\Mpz}{M_{\pi^0}}
\newcommand{\Meeg}{M_{ee\gamma}}
\newcommand{\Mgee}{M_{\gamma ee}}
\newcommand{\Mgg}{M_{\gamma\gamma}}
\newcommand{\Meeggg}{M_{ee\gamma\gamma\gamma}}

\newcommand{\Pgg}{\pi^0 \rightarrow \gamma \gamma}
\newcommand{\Peeg}{\pi^0 \rightarrow e^+ e^- \gamma}
\newcommand{\KLppp}{K_L \rightarrow \pi^0 \pi^+ \pi^-}
\newcommand{\KLpzee}{K_L \rightarrow \pi^0 e^+ e^-}
\newcommand{\KLpzpz}{K_L \rightarrow \pi^0 \pi^0}
\newcommand{\KLpzgg}{K_L \rightarrow \pi^0 \gamma \gamma}
\newcommand{\KLpzgee}{K_L \rightarrow \pi^0 \gamma e^+ e^-}

\newcommand{\KLeeg}{K_L \rightarrow e^+ e^- \gamma}
\newcommand{\KLpzpzd}{K_L \rightarrow \pi^0 \pi^0_D }
\newcommand{\KLpzpzpzd}{K_L \rightarrow \pi^0 \pi^0 \pi^0_D}
\newcommand{\KLppee}{K_L \rightarrow \pi^+ \pi^- e^+ e^-}

\newcommand{\Kpgg}{K^+ \rightarrow \pi^+ \gamma \gamma}

\begin{document}

\begin{frontmatter}
\journal{Physics Letters B}
\title{Experimental search for the decay mode 
$K_L \rightarrow \pi^0 \gamma e^+ e^-$ }
\author{K. Murakami},
\author{Y. Hemmi\thanksref{DIT}},
\author{H. Kurashige\thanksref{KOBE}},
\author{Y. Matono},
\author{T. Nomura\thanksref{CONTACT}},
\author{H. Sakamoto}, 
\author{N. Sasao},
\author{M. Suehiro},
\author{Y. Takeuchi}
\address
 {Department of Physics, Kyoto University, Kyoto 606-8502, Japan }
\author{Y. Fukushima}, 
\author{Y. Ikegami},
\author{T. T. Nakamura}, 
\author{T. Taniguchi}
\address
 {High Energy Accelerator Research Organization (KEK),
 Ibaraki 305-0801, Japan }
\author{M. Asai}
\address
 {Hiroshima Institute of Technology, Hiroshima 731-5193, Japan }
\thanks[DIT]
 {Present address: {\it Daido Institute of Technology, Aichi 457, Japan}}
\thanks[KOBE]
 {Present address: {\it Kobe University, Hyogo 657-8501, Japan}}
\thanks[CONTACT]
 {Contact person: nomurat@scphys.kyoto-u.ac.jp}

\begin{abstract}
We report on results of a search for the decay mode 
  $K_L \rightarrow \pi^0 \gamma e^+ e^-$ conducted 
  by the E162 experiment at KEK.
We observed no events and set a 90\% confidence level upper limit of 
  $Br(\KLpzgee)< 7.1 \times 10^{-7}$ for its branching ratio.
This is the first published experimental result on this decay mode.

\end{abstract}
\end{frontmatter}

\section{Introduction}
Chiral perturbation theory ($\chi$PT) is a very powerful tool 
  to describe various $K$ decays in which long distance 
  contributions are expected to dominate.
For example, the decay mode $\Kpgg$ has been observed recently,
  and compared with $\chi$PT~\cite{Kitching97}.
The measured branching ratio and $\pi^+$ momentum spectrum 
  are found to be consistent with the predictions, 
  after fitting one free parameter contained in the theory.
The neutral counterpart, $\KLpzgg$, is another decay mode, 
  where detailed studies have been performed.
In this case, the lowest order $O(p^4)$ calculation~\cite{EPR}
  does not reproduce the measured 
  branching ratio~\cite{Barr92,Papa91,Alavi99}.
Extending the calculation to the next-to-leading 
  order $O(p^6)$~\cite{CEP},
  and adding a vector meson 
  contribution~\cite{VMD}, 
  the prediction is now in good agreement with the branching ratio
  as well as distinct $M_{\gamma\gamma}$ spectrum.
We note that an effective coupling constant ($a_V$), 
  a free parameter in the vector meson contribution, 
  has been determined by the measurements 
  with experimental errors~\cite{Barr92,Alavi99}.
The as yet unobserved decay mode $\KLpzgee$\footnote{
    An observation of 18 events has been claimed by the KTEV 
    experiment in ICHEP 98 workshop}
  can provide another testing ground for $\chi$PT.
Since theoretical ingredients are same for 
  both $\KLpzgg$ and $\KLpzgee$ modes,
  a straightforward extension from $\KLpzgg$ provides 
  definite prediction for a branching ratio and 
  $\Mgee$ spectrum~\cite{Donoghue97}.
In particular, the branching ratio is calculated to be 
  $2.3 \times 10^{-8}$.
Thus experimental study of this mode is important to 
  test the theoretical framework of $\chi$PT.

Other interests in this decay mode stem from 
  its close relationship to the mode $\KLpzee$, 
  which has been of much attention as a possible channel to 
  observe direct $CP$-violation.
First, $\KLpzgee$ is expected to have a much larger branching ratio
  than $\KLpzee$, and hence can be an experimental background
  in a soft photon region.
Second, there also exists a $CP$-conserving amplitude in 
  $\KLpzee$ via two-photon intermediate states; 
  this can be in principle determined from a detail analysis
  of $\KLpzgg$~\cite{Donoghue95}.
Further understanding of the $\KLpzgg$ amplitude, 
  which can be checked by the $\KLpzgee$ mode, 
  is thus essential.

In this article, we report on an experimental
  search for the decay mode $\KLpzgee$ 
  conducted with a proton synchrotron at High Energy 
  Accelerator Research Organization (KEK).

\section{Experimental setup}
The data for the $\KLpzgee$ mode were recorded 
  simultaneously with the experiment 
  which has established a new decay mode 
  $\KLppee$~\cite{Takeuchi98}.
Since the experimental set-up was described already  
  in Ref.~\cite{Takeuchi98,Nomura97},
  it is briefly mentioned here for convenience.

The $K_L$ beam was produced by focusing 12-GeV/c 
 primary protons onto a 60-mm-long copper target. 
Its divergence was
 $\pm$4~mrad horizontally and $\pm$20~mrad vertically
  after a series of collimators embedded in magnets.
The set-up started with a 4-m-long decay volume.
It was followed by a charged particle spectrometer
  consisting of four sets of drift chambers and 
  an analyzing magnet with an average horizontal
  momentum kick of 136~MeV/c.
A threshold Cherenkov counter (GC) with pure N$_2$ gas at 1 atm 
 was placed inside of the magnet gap to identify electrons.
For the present decay mode, we obtained
  an average electron efficiency of {94\%} with 
  a pion-rejection factor of {350}
  by adjusting software cuts in the off-line analysis.
A pure CsI electromagnetic calorimeter
  played a crucial role in this analysis.
It was located at the downstream end of the spectrometer,
  and consisted of 540 crystal blocks, each being
  70~mm by 70~mm in cross section and 300~mm ($\sim$ 16X$_{0}$)
  in length.
It was configured into two banks of 15 (horizontal) 
  $\times $ 18 (vertical) matrix.
Its energy and position resolutions were found to be approximately 3\%
  and 7~mm for 1-GeV electrons, respectively.
The trigger for the present mode was designed to select
 events with $\geq$~2 electron-like tracks and 
  $\geq$~3 cluster candidates in the calorimeter.
It was formed with information from GC and CsI together with 
 trigger scintillator hodoscopes interspersed between the
 chambers and calorimeter.

\section{General event selection}
In reconstructing $\KLpzgee$ events, major backgrounds are 
  expected come from both
  $\KLpzpzd$ and $\KLpzpzpzd$ modes, where $\pi^0_{D}$ denotes the Dalitz 
  decay $\pi^0 \rightarrow e^+ e^- \gamma$.
Among them, $\KLpzpzd$  is an intrinsic background and can not be removed.
It is instead used as a normalization mode by positively identifying
  the Dalitz decay.
The $\KLpzpzpzd$ mode may become background when two
  (or more) photons fuse into one in the calorimeter,
  and/or its vertex is reconstructed incorrectly
  to give false invariant masses.
Care was taken in this analysis to enhance 
  position resolution of decay vertex and 
  purity of photon clusters.
This effort was found useful also to 
  reject backgrounds originating from external conversions
  and hadron (mostly neutron) interactions.
  
In the actual off-line analysis, we first selected events containing 
  two tracks with a common vertex in the beam      
  region inside the decay volume.
We then requested events to have $\geq$~5 clusters in the
  calorimeter.
Here the cluster was defined as $3\times 3$ CsI blocks around the 
  local maximum with the total
  energy deposit greater than 200~MeV 
  ($\pm$ 3.5~nsec timing window).
Particle species were then determined.
A charged track which could project onto a CsI cluster
  was called a matched track.
An electron (or positron) was identified as a matched track
  with  0.9 $\leq$ E/p $\leq$ 1.1,
  where E was  an energy measured by the calorimeter 
  and {p} was a momentum determined by the spectrometer, 
  respectively.
Also GC hits in the corresponding cells were requested.
Clusters which did not match with any charged tracks
  were treated as photon candidates.

Event topology was then checked.
We requested events to contain
  exactly one $e^+e^-$-pair and three photon candidates, 
  consistent with the $\KLpzgee$ topology.
No additional activities, 
  such as an extra track, GC hit, or cluster
  with an energy above $\sim$60~MeV,
  was allowed in the detector\footnote{
    The exception was bremsstrahlung photons. 
    When e$^{\pm}$ track segment upstream of the magnet could be
    projected onto a neutral cluster with an energy below 400~MeV, 
    we simply ignored this activity.}.
The probability of over-veto was estimated  
  using $\KLpzpzd$ and $\KLeeg$ reconstructed events, 
  and was found to be about 12\% 
  (common to both $\KLpzgee$ and $\KLpzpzd$ modes). 

Two quality cuts were imposed at this stage.
One was a cluster shape cut; 
  it examined mainly a cluster's transverse energy profile,
  and checked whether or not 
  it was consistent with a single photon.
We used a large sample of $\KLppp$ reconstructed events
  to characterize actual photon showers,
  and found the cut efficiency 
  to be 95\% per a single photon cluster.
The other quality cut was imposed on an $e^{+}e^{-}$-track opening angle 
  ($\theta_{ee}$).
Larger opening angle resulted generally in
  better vertex position resolution, which
  in turn led to better invariant mass resolution and
  background rejection.
We employed a Monte Carlo simulation to study 
  background rejection power, especially for the
  $\KLpzpzpzd$ mode, and determined to demand $\theta_{ee}>$~20~mrad.

As a final step in the general event selection, 
  we imposed two loose kinematical cuts to reduce sample size.
They were $\Meeggg>$~400~MeV/c$^2$
  and $\theta^{2}<$~100~mrad$^{2}$, 
  where $\theta$ represents the angle of the reconstructed
  $ee \gamma \gamma \gamma$ momentum with respect to the 
  line connecting the target and vertex.

\section{Kinematical reconstruction}
\begin{figure*}[ht]
\centerline{
\epsfysize=9cm
\epsfbox{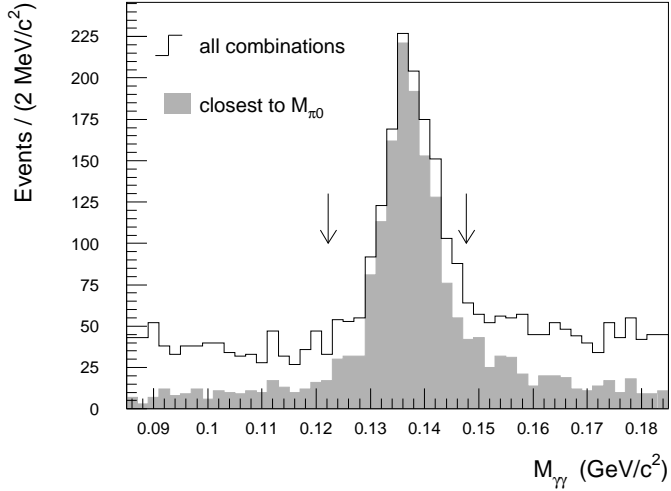}}
\vspace*{-0.8cm}
\caption{ 
  The invariant mass distribution of $\gamma\gamma$.
  The solid-line histogram shows the distribution for all 
  $\gamma\gamma$ combinations  while the shaded one 
  for the combination closest to $\Mpz$.
  The arrows indicate the cut position.
}
\label{fig:gamma-gamma}
\end{figure*}

Having selected candidate events, we scrutinized each event
  from the viewpoint of kinematical variables.
First of all, we calculated $\gamma \gamma$ invariant mass
  for three possible combinations.
The solid-line histogram shown in Fig.\ref{fig:gamma-gamma}
  is the distribution for all combinations
  while the shaded one for the combination closest to $\Mpz$.
We retained all combinations which satisfied 
  $|\Mgg - \Mpz|<$~2.5$\sigma_{\pi^0}$\footnote{
    This condition was determined in such a way that the number of 
    final $\KLpzpzpzd$ background events,
    estimated by a Monte Carlo simulation,
    became less than one inside the final signal box 
    (see below).},
  where  $\sigma_{\pi^0}$ 
  being the observed mass resolution of 5.1~MeV/c$^2$.

\begin{figure*}[ht]
\centerline{
\epsfysize=9cm
\epsfbox{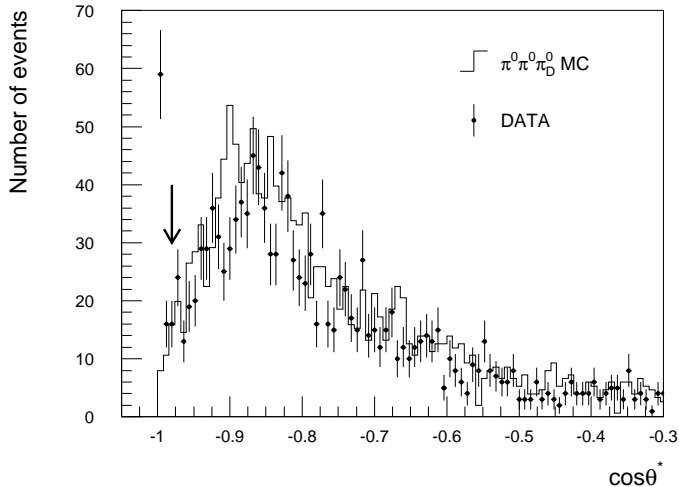}}
\vspace*{-0.8cm}
\caption{   
  The distribution of $cos \theta^*$.
  The dots with error bars show the experimental data,
  and the histogram is Monte Carlo data for $\KLpzpzpzd$ , 
  in which the flux was normalized by the observed $\KLpzpzd$ events.
  The cut position ($cos\theta^*<$~-0.98) is indicated by the arrow.
}
\label{fig:cos-star}
\end{figure*}

We then calculated the angle ($\theta^{\ast}$) 
  between reconstructed $\pi^{0}$ and $ee \gamma$ momentum vectors
  in the $K_L$ rest frame\footnote{
    Actually this frame was obtained with the Lorenz boost,
    along the line connecting the target and vertex, 
    defined by the velocity of $K_L$ with
    an observed $ee\gamma\gamma\gamma$ energy.}.
Events originating from $\KLpzgee$ or $\KLpzpzd$ should satisfy
  $\cos \theta^{\ast}=-1$ in this frame.
The dots with error bars in Fig.\ref{fig:cos-star} 
  show the $\cos \theta^{\ast}$ 
  distribution; a clear peak of events 
  at $\cos \theta^{\ast}=-1$ can be seen.
The histogram in Fig.\ref{fig:cos-star} is Monte Carlo data
  for the $\KLpzpzpzd$ mode, in which the flux was normalized 
  by the observed $\KLpzpzd$ events (see below).
To select signals, we requested events to satisfy 
  a collinearity cut, $\cos \theta^{\ast}<-0.98$, as shown by 
  the arrow in  Fig.\ref{fig:cos-star}.

We now identify $\KLpzpzd$ events.
If more than one $\gamma \gamma$ combination within an event 
  satisfied the $M_{\pi^{0}}$ cut, 
  we selected the one for which the quantity,
   \[
   \chi^2_p =
     \left(
     \frac{M_{ee\gamma}-M_{\pi^0}}{\sigma_{\pi^0_D}} \right)^2 +
     \left(
     \frac{M_{\gamma\gamma}-M_{\pi^0}}{\sigma_{\pi^0}} \right)^2, 
  \]
  became minimum.
Here $\sigma_{\pi^{0}_D}$ is the observed $\pi^{0}$
  mass resolution for the Dalitz decay mode 
  (see below for the actual value).

\begin{figure*}[ht]
\centerline{
\epsfysize=13cm
\epsfbox{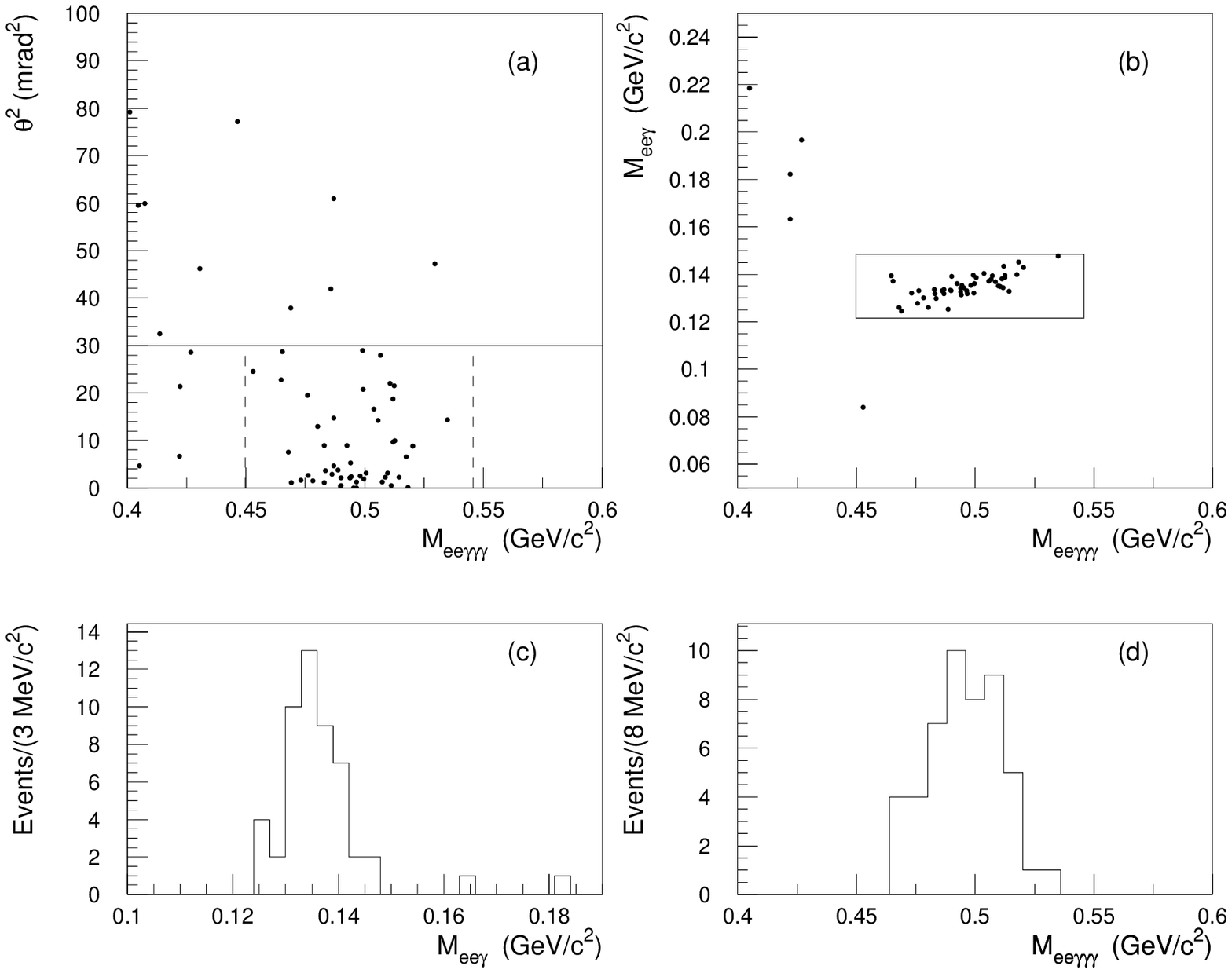}}
\vspace*{-0.8cm}
\caption{ 
(a) The $\Meeggg$ vs $\theta^2$ scatter plot
  of the $\KLpzpzd$ candidate events.
(b) The scatter plot of $\Meeggg$ vs $\Meeg$
  for the events with $\theta^2<$~30~mrad$^2$.
  The box indicates the signal region.
(c) The $\Meeg$ projection of the events in (b).
(d) The $\Meeggg$ projection of the events in (b)
  with the $\pi^0_D$ mass cut (see text).
}
\label{fig:p0p0d}
\end{figure*}

Fig.\ref{fig:p0p0d}(a) shows a scatter plot of  
  $\Meeggg$ vs $\theta^{2}$ after selecting the 
  $\gamma \gamma$ combination with $\chi^2_p$.
To select $\KLpzpzd$ events further, $\theta^{2}<$~30~mrad$^{2}$
  was demanded.
Fig.\ref{fig:p0p0d}(b) shows a scatter plot of 
  $\Meeggg$ vs $\Meeg$ after this cut.
A clear cluster of  $\KLpzpzd$ events can be seen in the
   expected region of $\Meeg=\Mpz$ and $\Meeggg=\MKL$.
Fig.\ref{fig:p0p0d}(c) is the projection onto the $\Meeg$ axis, 
  and Fig.\ref{fig:p0p0d}(d) is onto the $\Meeggg$ axis 
  with a $\pi^0_D$ mass cut (see below).
From these projections, we found  
  $\pi^0_D$ and $K_L$ mass resolutions
  to be $\sigma_{\pi^0_D} \simeq$ 4.5~MeV/c$^2$ and 
  $\sigma_{K_L} \simeq$ 16~MeV/c$^2$, respectively.
Our final signal box,
  shown by the rectangle in  Fig.\ref{fig:p0p0d}(b),
  was defined by
   $|\Meeg-\Mpz|<~$3$\sigma_{\pi^0_D}$ 
   and
   $|\Meeggg-\MKL|<$~3$\sigma_{K_L}$.
After all the cuts, 49 events remained.
We estimated the number of $\KLpzpzpzd$ backgrounds in the signal region
  by a Monte Carlo simulation, and found to be less than one.

\begin{figure*}[ht]
\centerline{
\epsfysize=90mm
\epsfbox{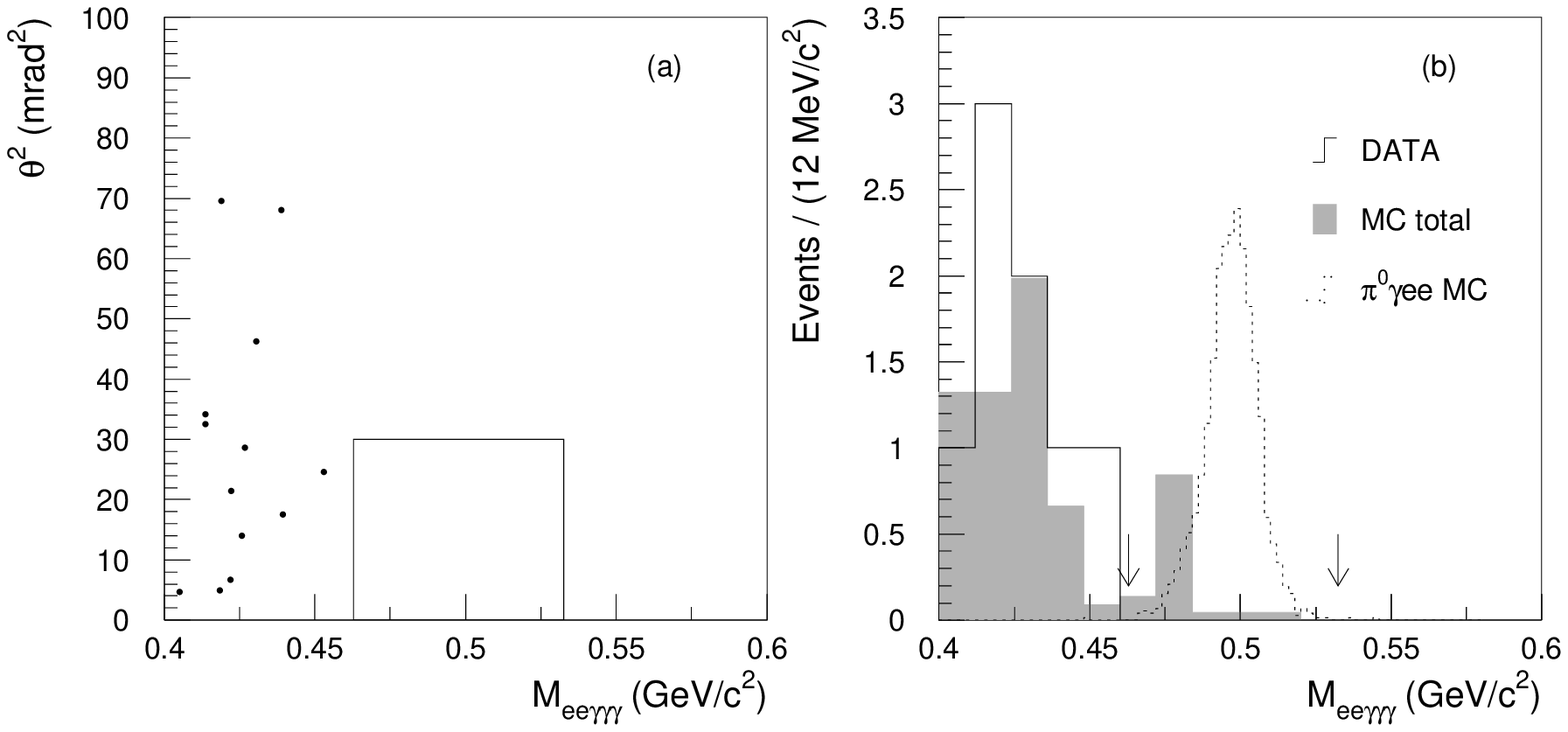}}
\vspace*{-0.8cm}
\caption{
(a) The scatter plot of $\Meeggg$ vs $\theta^2$ for $\KLpzgee$ 
  candidate events. 
There are no events left inside the signal box.
(b) The $\Meeggg$ projection of the events with $\theta^2<$~30~mrad$^2$ 
  for the data (solid-line), 
  Monte Carlo (shaded, sum of $\KLpzpzd$ and $\KLpzpzpzd$) and 
  $\KLpzgee$ Monte Carlo (dotted, arbitrary scale).
The arrows show the signal region.
}
\label{fig:p0p0p0d}
\end{figure*}

We are now in a position to look for the $\KLpzgee$ mode.
First we rejected $\KLpzpzd$ events; 
  if an event satisfied both 
  $|\Mgg-\Mpz|<$~5$\sigma_{\pi^0}$ 
  and
  $|\Meeg-\Mpz|<$~5$\sigma_{\pi^0_D}$
  for any $\gamma$ combinations,
  then the event was discarded.
Note that we employed the looser kinematical cut of 
  5$\sigma_{\pi^0}$($\sigma_{\pi^0_D}$)
  to exclude possible $\KLpzpzd$ events.
Then we rejected $\KLpzpzpzd$ backgrounds.
In this case, an event containing $\pi^0_D$ whose
  transverse momentum was consistent with  
  $\KLpzpzpzd$ (p$_{t}<$~139~MeV/c) was excluded.  
The remaining events are shown in a scatter plot of 
  $\Meeggg$ vs  $\theta^{2}$ in Fig.\ref{fig:p0p0p0d}(a).
We are left with no events inside our signal box defined by
  $\theta^{2}<$~30~mrad$^2$
  and
  $|\Meeggg-\MKL|<$~3$\sigma^{\ast}_{K_{L}}$\footnote{
    The actual value of $\sigma^{\ast}_{K_L}$ was 11.6~MeV/c$^2$.
    This was the value of the Monte Carlo mass resolution
    (8~MeV/c$^2$) for $\KLpzgee$ 
    times the ratio of the observed ($16 \pm 2.6$~MeV/c$^2$) 
    to Monte Carlo (11~MeV/c$^2$) resolutions for $\KLpzpzd$.
    }.
The background events still remaining in the low mass region of 
  $\Meeggg <$~460~MeV/c$^2$ were found to originate 
  mostly from the $\KLpzpzpzd$ mode.
The projection of events with  $\theta^{2}<$~30~mrad$^2$ 
  onto the $\Meeggg$ axis is shown in Fig.\ref{fig:p0p0p0d}(b).
The solid-line histogram is for the data,
  the shaded one for Monte Carlo events 
  (sum of $\KLpzpzd$ and $\KLpzpzpzd$)
  and the dotted one for $\KLpzgee$  Monte Carlo events.
From the Monte Carlo simulation,
   whose flux was normalized by the observed $\KLpzpzd$ signals,
  we expected 1.1 backgrounds (0.45 from $\KLpzpzd$ and 
  0.66 from $\KLpzpzpzd$) to remain in the signal box.

%

\section{Results}
The branching ratio is calculated by 
\[
\begin{array}{lcl}
Br(\KLpzgee) & = & Br(\KLpzpzd) \\
& \times & \displaystyle
\frac{A(\pi^0\pi^0_D)}{A(\pi^0\gamma e^+e^-)} 
  \cdot  \frac{\eta(\pi^0\pi^0_D)}{\eta(\pi^0\gamma e^+e^-)} 
  \cdot  \frac{N(\pi^0\gamma e^+e^-)}{N(\pi^0\pi^0_D)},
\end{array}
\]
where $A$, $\eta$ and $N$ denote acceptance, efficiency and 
  observed number of events, respectively.
The detector acceptances depend on the matrix elements: 
  we employed a theoretical model given by Ref.~\cite{Donoghue97}
  for $\KLpzgee$ and Kroll-Wada spectrum for 
  the Dalitz decay~\cite{KW55}.
The actual acceptances (including the hardware trigger conditions)
were determined by Monte Carlo simulations
  and were found to be $2.34 \times 10^{-4}$ for $\KLpzgee$ 
  and $1.38 \times 10^{-4}$ for $\KLpzpzd$, respectively.
The detector efficiency was found to be 
  practically same for both modes.
However, some of the quality and kinematical cuts caused 
  efficiency differences.

Since the opening angle of $e^+e^-$-tracks was different 
  for the two modes, it caused efficiency difference
  in both the vertex reconstruction and the track opening 
  angle cut.
These efficiencies were found to be
  70\% (vertex) and 89\% ($\theta_{ee}$) for $\KLpzgee$ mode 
  while 62\% and 81\% for $\KLpzpzd$ mode, respectively.
Next, the collinearity ($\theta^\ast$) cut  produced 
  25\% inefficiency for the signal mode while negligibly small 
  loss for the normalization mode.
The $\pi^0_D$ mass cut was unique to $\KLpzpzd$;
  its efficiency was found to be 96\% (3$\sigma$ cut).
For the signal mode we rejected  $\KLpzpzd$ events;
 it caused 9\% inefficiency at the specific $ee\gamma$ invariant mass region.
We also rejected $\KLpzpzpzd$ events with the $\pi^{0}_{D}$ inclusive veto;
 it resulted in 7\% inefficiency.
Combining other efficiencies together, we found 
 the final acceptance and efficiency ratio 
 $A\eta(\pi^0\pi^0_D)/A\eta(\pi^0\gamma e^+e^-)$ 
 to be 0.670. 

Inserting the known branching ratios~\cite{PDG} into 
  $Br(\KLpzpzd) = 2 \times Br(\KLpzpz) \times Br(\Pgg) \times Br(\Peeg)$,
  the single event sensitivity was obtained 
  to be $(3.03 \pm 0.43) \times 10^{-7}$, 
  where the error represents the statistical uncertainty.
The upper limit on the branching ratio 
  was determined to be
\[
     Br(\KLpzgee )< 7.1 \times 10^{-7}
     \qquad \mbox{(90\% C.L.)},
\]
  in which the statistical error of the normalization mode 
  was taken into account~\cite{Cousins92}.
 
In summary, we performed an experimental search for the $\KLpzgee$ mode.
We observed no events and set a 90\% confidence level upper limit of 
  $Br(\KLpzgee )< 7.1 \times 10^{-7}$ for its branching ratio.
This is the first published experimental result on this decay mode.

\begin{ack}
We wish to thank Professors H. Sugawara, S. Yamada,  S. Iwata,
 K. Nakai, and K. Nakamura
 for their support and encouragement. 
We also acknowledge the support from the operating crew of the
 Proton Synchrotron, 
 the members of Beam Channel group, Computing Center and 
 Mechanical Engineering Center 
 at KEK. 
Y.T, Y.M and M.S acknowledge receipt of Research Fellowships
 of the Japan Society for the Promotion of Science for Young Scientists.
\end{ack}

%
%

\end{document}